\documentclass{aastex631}
\hypersetup{linkcolor=red,citecolor=green,filecolor=cyan,urlcolor=magenta}

\graphicspath{{./}{figures/}}
\usepackage{amsmath}
\usepackage{mwe}
\usepackage{hyperref}

\revised{\today}

\submitjournal{ApJ. Lett.}

\shorttitle{J0348+0432}
\shortauthors{Saffer et al.}

\begin{document}

\title{A Lower Mass Estimate for PSR J0348+0432 Based on CHIME/Pulsar Precision Timing}
\correspondingauthor{Alexander Saffer}
\email{asaffer@nrao.edu}

\author[0000-0001-7832-9066]{Alexander Saffer}
\affiliation{National Radio Astronomy Observatory, 520 Edgemont Rd., Charlottesville, VA 22903, USA}
\affiliation{Department of Physics and Astronomy, West Virginia University, P.O. Box 6315, Morgantown, WV 26506, USA}
\affiliation{Center for Gravitational Waves and Cosmology, West Virginia University, Chestnut Ridge Research Building, Morgantown, WV 26505, USA}

\author[0000-0001-8384-5049]{Emmanuel Fonseca}
\affiliation{Department of Physics and Astronomy, West Virginia University, P.O. Box 6315, Morgantown, WV 26506, USA}
\affiliation{Center for Gravitational Waves and Cosmology, West Virginia University, Chestnut Ridge Research Building, Morgantown, WV 26505, USA}

\author[0000-0001-5799-9714]{Scott Ransom}
\affiliation{National Radio Astronomy Observatory, 520 Edgemont Rd., Charlottesville, VA 22903, USA}

\author[0000-0001-9784-8670]{Ingrid Stairs}
\affiliation{Dept. of Physics and Astronomy, University of British Columbia, 6224 Agricultural Road, Vancouver, BC V6T 1Z1 Canada}

\author[0000-0001-5229-7430]{Ryan Lynch}
\affiliation{National Radio Astronomy Observatory, 520 Edgemont Rd., Charlottesville, VA 22903, USA}

\author[0000-0003-1884-348X]{Deborah Good}
\affiliation{Department of Physics and Astronomy, University of Montana, 32 Campus Drive, Missoula, MT 59812}

\author[0000-0002-4279-6946]{Kiyoshi W. Masui}
\affiliation{MIT Kavli Institute for Astrophysics and Space Research, Massachusetts Institute of Technology, 77 Massachusetts Ave, Cambridge, MA
02139, USA}
\affiliation{Department of Physics, Massachusetts Institute of Technology, 77 Massachusetts Ave, Cambridge, MA 02139, USA}

\author[0000-0002-2885-8485]{James W. McKee}
\affiliation{Department of Physics and Astronomy, Union College, Schenectady, NY 12308, USA}

\author[0000-0001-8845-1225]{Bradley W. Meyers}
\affiliation{International Centre for Radio Astronomy Research, Curtin University, Bentley, WA 6102, Australia}

\author[0009-0008-7264-1778]{Swarali Shivraj Patil}
\affiliation{Department of Physics and Astronomy, West Virginia University, P.O. Box 6315, Morgantown, WV 26506, USA}
\affiliation{Center for Gravitational Waves and Cosmology, West Virginia University, Chestnut Ridge Research Building, Morgantown, WV 26505, USA}

\author[0000-0001-7509-0117]{Chia Min Tan}
\affiliation{International Centre for Radio Astronomy Research, Curtin University, Bentley, WA 6102, Australia}

\begin{abstract}
    The binary pulsar J0348+0432 was previously shown to have a mass of approximately 2\,${\rm M_\odot}$, based on the combination of radial-velocity and model-dependent mass parameters derived from high-resolution optical spectroscopy of its white-dwarf companion.
    We present follow-up timing observations that combine archival observations with data acquired by the Canadian Hydrogen Intensity Mapping Experiment (CHIME) pulsar instrument. 
    We find that the inclusion of CHIME/Pulsar data yields an improved measurement of general-relativistic orbital decay in the system that falls within 1.2 $\sigma$ of the original values published by~\citet{Antoniadis:2013pzd} while being roughly 6 times more precise due to the extended baseline.
    When we combine this new orbital evolution rate with the mass ratio determined from optical spectroscopy, we determine a pulsar mass of 1.806(37)\,${\rm M_\odot}$. 
    For the first time for this pulsar, timing alone significantly constrains the pulsar mass.
    We explain why the new mass for the pulsar is $10\%$ lower and discuss how the mis-modeling of the initial observations of the white dwarf companion likely led to an inaccurate determination of the pulsar mass. 
\end{abstract}


\section{Introduction} \label{sec:intro}
Neutron star (NS) masses have been of keen interest to astronomers and physicists over the past century.
Ultra-dense nuclear matter cannot be probed in a terrestrial laboratory, but is accessible via observations of neutron stars, with relevant properties being the radius and mass.
Various equation of state (EoS) models have suggested the maximum mass of a NS to be somewhere between 2.1 and 2.5\,${\rm M_\odot}$.
Theoretical limits on the upper NS mass have been presented in ~\citet{Lattimer:2000nx,Chamel:2013efa,Ozel:2016oaf}.

Measurements of large NS masses can constrain or eliminate ranges of possible  EoS.
Since a greater observed NS mass would yield a limit on dense nuclear matter, a 2\,${\rm M_\odot}$ NS has been seen as a promising milestone in the study of compact objects.
Of course, the observation of the mass will give insight into the composition and EoS for these compact objects, allowing scientists to better understand the fundamentals of ultra-dense nuclear matter~\citep{Shao:2022koz,MUSES:2023hyz}.

There are several NSs whose masses have approached (or exceeded) 2\,${\rm M_\odot}$. In 2010,~\citet{Demorest:2010bx} made use of the observed Shapiro delay of PSR J1614$-$2230 to claim a mass bound of $1.97 \pm 0.04$\,${\rm M_\odot}$, making it the first NS to have a mass possibly greater than 2\,${\rm M_\odot}$ 
\footnote{Recently,~\citet{NANOGrav:2023hde} placed a new bound on this mass of J1614$-$2230 to be $1.937 \pm 0.014$\,${\rm M_\odot}$}.
Several years later,~\citet{NANOGrav:2019jur} and~\citet{Fonseca:2021wxt} used the same method to determine the mass of another NS, PSR J0740+6620, at $2.08 \pm 0.07$\,${\rm M_\odot}$.
These studies relied on the well established relativistic Shapiro time delay \citep{Shapiro64}.
For orbits seen nearly edge on, the radio signals from the pulsar will be slightly delayed via the gravitational field of the companion due to the increased light travel time caused by following the geodesic past the companion star..
Using this, the theory of General Relativity can provide direct measurements of the mass of the pulsar's companion as well as the sine of the orbital inclination angle.
These quantities can be combined with the Keplerian mass function, which is determined extremely precisely via pulsar timing, to obtain a measurement of the neutron star mass.
Some systems permit the measurement of other relativistic parameters, which can lead to similar or redundant mass constraints~\citep{1989ApJ...345..434T}. 

In addition to the Shapiro delay method, astronomers may combine pulsar timing with optical spectroscopy of a companion to determine the orbital characteristics (and hence the mass) of a NS (e.g.~\citet{Linares:2018ppq,Romani:2021xmb,Romani:2022jhd,Sen:2024xbs}).
This method is analogous to the double-lined spectroscopic method used to study binaries that observes how the spectral lines of a NS companion, typically a white dwarf (WD), shift with orbital phase.

When combined with the velocity of the pulsar from timing observations, the spectroscopic data can be used to determine the mass ratio $q=M_\mathrm{NS}/M_\mathrm{WD}$ of the system.
Additionally, the width of the Balmer lines can give insight into the WD's gravitational acceleration near the surface of the star.
Depending on the WD modeling used, these factors can all be used to determine the mass of the WD, and then that of the NS via the mass ratio $q$.
\cite{Antoniadis:2013pzd} used this method when investigating J0348+0432, a compact NS-WD binary.
In their work, the spectroscopic results led to a mass ratio of $q=11.70 \pm 0.04$.
Then, via WD modeling, they determined the mass of the WD to be $0.172 \pm 0.003$\,${\rm M_\odot}$ and the NS mass to be $2.01 \pm 0.04$\,${\rm M_\odot}$.
Additionally the original study was able to place a limit on the binary orbital evolution rate, a quantity which can be used to test General Relativity by assuming orbital energy loss is due to gravitational radiation.

The work contained within this Letter follows-up the work by \citet{Antoniadis:2013pzd} with new pulsar timing observations with the Canadian Hydrogen Intensity Mapping Experiment (CHIME) telescope~\citep{Amiri_2021,CHIME:2022dwe}.
We extend the timing baseline by a factor of 4 to place updates on the mass and binary evolution parameters for PSR J0348+0432.

\section{Observations \& Data Reduction} \label{sec:obs}
Our study is based on the combination and analysis of two distinct data sets: the timing data published by \cite{Antoniadis:2013pzd}; and new timing data acquired with the CHIME telescope taken from July 01, 2020 to January 30, 2023. Details of the archival observations and their reduction are provided by \cite{Antoniadis:2013pzd}. PSR J0348+0432 is observed by the CHIME telescope on a near-daily basis. The CHIME telescope uses a dedicated instrument for producing coherently dedipsersed pulse profiles across the 400$-$800 MHz range, which we hereafter refer to as the CHIME/Pulsar backend \citep{chimepulsar21}. 

The CHIME/Pulsar backend receives a stream of voltage data with frequency and time resolutions of 0.390625 MHz and 2.56 $\mu$s, respectively; the instrument then performs real-time coherent dedispersion and folding of the channelized timeseries based on an existing model of pulsar rotation, using the {\tt dspsr} software package \citep{vb11}. For all observations of PSR J0348+0432, we used the timing solution that is publicly available on the Australia Telescope National Facility (ATNF) Pulsar Catalogue to produce initial data products \citep{atnfcatalog}. These data consist of coherently dedispersed profiles formed over 1,024 channels, 256 profile bins, and 10-second ``subintegrations" that span the transit time of the pulsar, which is typically $\sim$13 minutes at the declination of the pulsar.

These raw, folded profile data were cleaned and downsampled using the {\tt clfd}~\citep{2019MNRAS.483.3673M,2023ascl.soft10008M} and {\tt psrchive}~\citep{2004PASA...21..302H,2010PASA...27..104V,2011ascl.soft05014V,2012AR&T....9..237V} software packages. For each observing epoch, we initially produced pulse profiles averaged across integration time but resolved into 8 frequency channels over the 400--800 MHz range. However, we found that the CHIME observations consistently yielded low S/N per epoch, even after full integration of the data in all dimensions. We therefore instead produced pulse profiles integrated over time and frequency, in order to maximize TOA precision.

\begin{figure}
    \centering
    \includegraphics[width=0.65\textwidth]{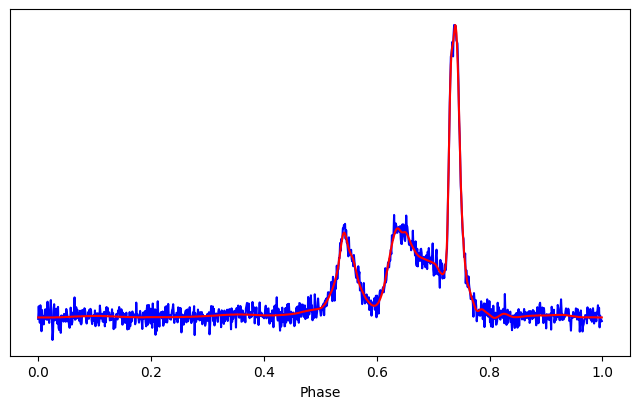}
    \caption{Pulse profile showing the relative flux for J0348+0432.  Both the averaged (blue) and smoothed (red) profiles are shown.}
    \label{fig:Profile}
\end{figure}

Once the cleaned products were formed, we combined all available CHIME data to generate an accurate, de-noised template for the pulse profile. The results of this analysis are shown in Fig.~\ref{fig:Profile}.
This combination was achieved by temporally aligning the 517 viable days of CHIME data into a single data file based on the publicly available timing solution on the ATNF pulsar database; we then fully downsampled this combined data set in both integration time and frequency before smoothing the resultant profile using the wavelet-based denoising {\tt psrsmooth} algorithm in {\tt psrchive}.
We subsequently used the template profile to compute TOAs for all CHIME observations using the {\tt pat} utility within {\tt psrchive}.
This command makes use of a Fourier Phase Gradient algorithm detailed in~\citet{1992RSPTA.341..117T}.

Lastly, we make use of two white-noise parameters (EFAC \& EQUAD) and fit for them with {\tt PINT}\footnote{\href{https://nanograv-pint.readthedocs.io/en/latest/}{https://nanograv-pint.readthedocs.io/en/latest/}}~\citep{2019ascl.soft02007L,Luo:2020ksx,Susobhanan:2024gzf}.
The EFAC parameter applies a constant scale factor to all raw TOAs, while the EQUAD adds an uncertainty term in quadrature to the TOAs.
Each telescope/back-end combination had its own white noise parameters, and {\tt PINT} can fit for these simultaneously.
Following this procedure, we obtain a timing model that yields a best fit reduced $\chi^2$ value of 1.0022.
\begin{figure}
    \centering    \includegraphics[width=0.7\textwidth]{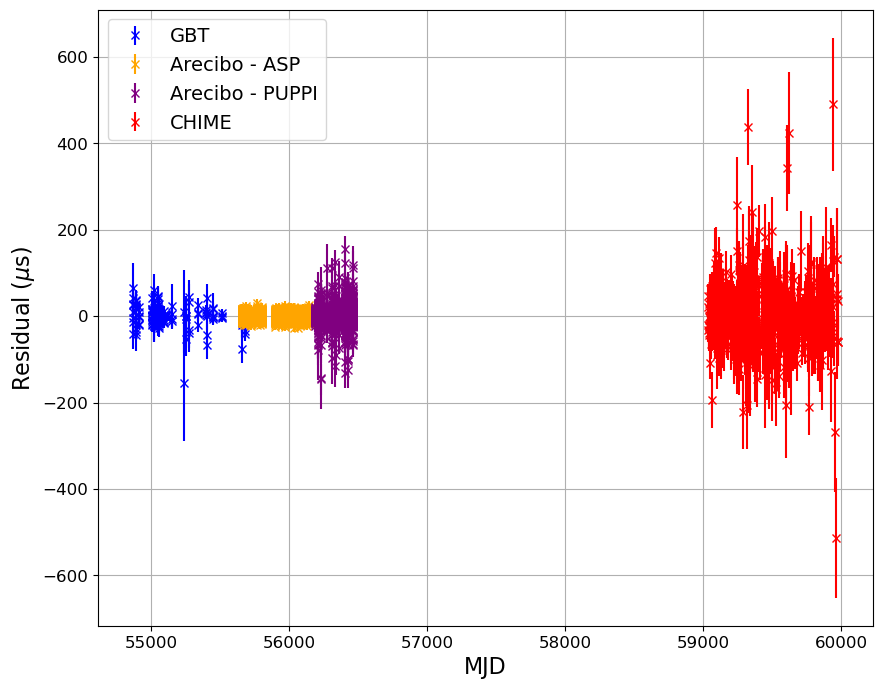}
    \caption{Post-fit timing residuals for J0348+0432 with the combined archival and newer CHIME data.  Original data from GBT comes from a drift scan survey~\citep{Lynch:2012vv}.  The Arecibo data is parsed into two groups differentiated by the backend.}
    \label{fig:Residuals}
\end{figure}
\begin{figure}
    \centering    \includegraphics[width=\textwidth]{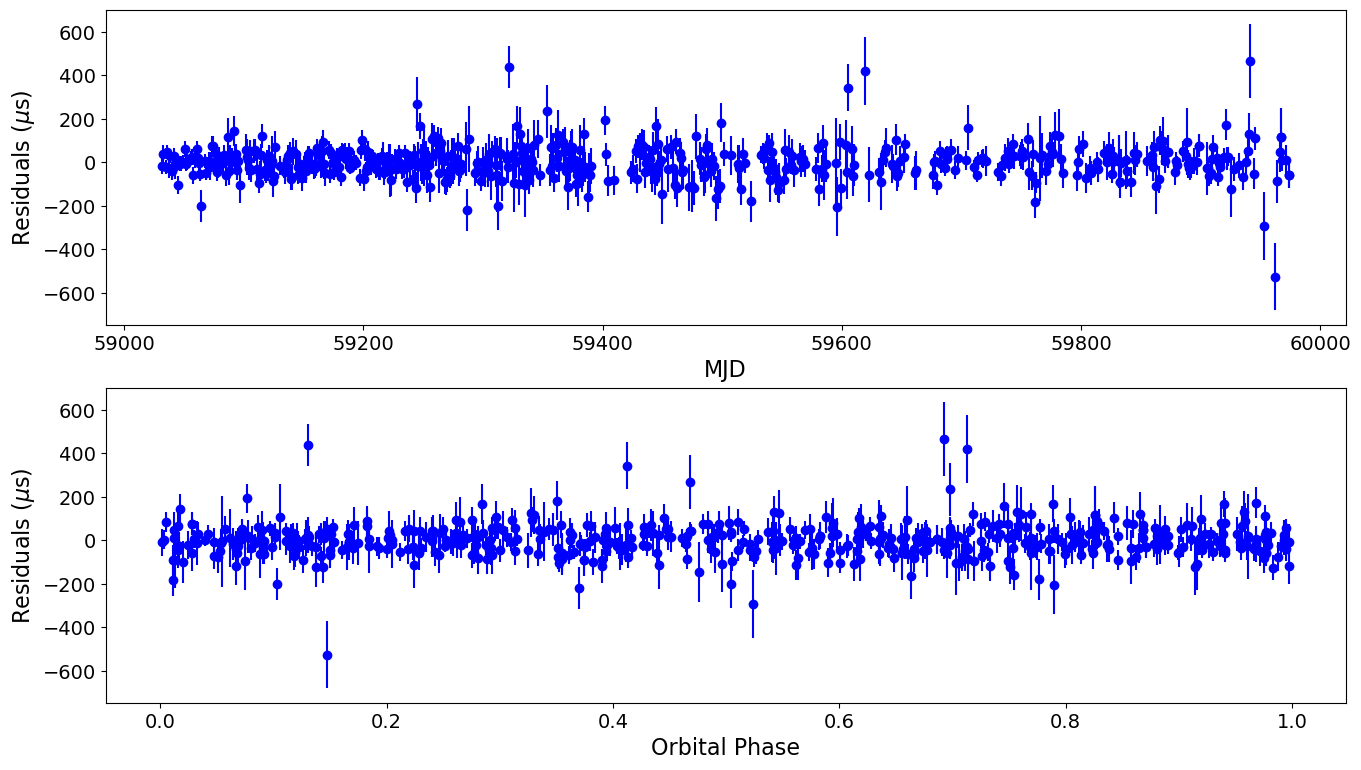}
    \caption{CHIME/Pulsar only residuals plotted as a function of time and orbital phase.  The analysis here yielded a best fit reduced $\chi^2$ value of 1.0022.}
    \label{fig:ChimeOnly}
\end{figure}

\subsection{Evaluating Dispersion Variations with CHIME Data}
\label{subsec:DM_CHIME}
As described above, we created a frequency-averaged TOA data set from CHIME/Pulsar data in order to maximize timing precision for evaluating orbital variations over time.
However, the low-frequency nature of the CHIME/Pulsar data set may nonetheless contain enough significance to constrain variations in pulse dispersion.
We explored the dispersion measure variation with both a DMX and wideband timing measurements.
When we made use of the DMX analysis later in our work, our data was frequency scrunched to 8 channels.
The number of channels was based on having a large enough selection of frequency channels, while still maintaining a reasonable signal to noise.
We used the {\tt PulsePortraiture} framework for simultanously estimating a ``wideband" TOA and dispersion measure DM = $\int_0^d n_e(l)dl$ from the broadband CHIME/Pulsar data set. These DM data can then be searched for, e.g., orbital modulations in DM. We did not use these wideband TOAs in the analysis of the combined TOA data set as archival data were produced using the ``narrowband" methods described above, and wideband data are analyzed differently by available pulse-timing software (see Sec.~\ref{sec:mLossandDM}).

\section{Results}
The results of our timing analysis are shown in Table~\ref{tab:TimingResults}.
In total, we have 9357 TOAs: 8676 from Arecibo, 164 from the GBT, and 517 from CHIME/Pulsar.
We process these data with the {\tt PINT} software package to derive the timing parameters for J0348+0432 and its white dwarf companion.
Additionally, we use the the Solar System ephemeris DE440~\citep{Park_2021} to correct the topocentric TOAs used in our analysis to the Solar System barycenter and the clock timescale BIPM2022.

\begin{figure}
    \centering
    \begin{minipage}{0.49\textwidth}
        \centering
        \includegraphics[width=0.75\textwidth]{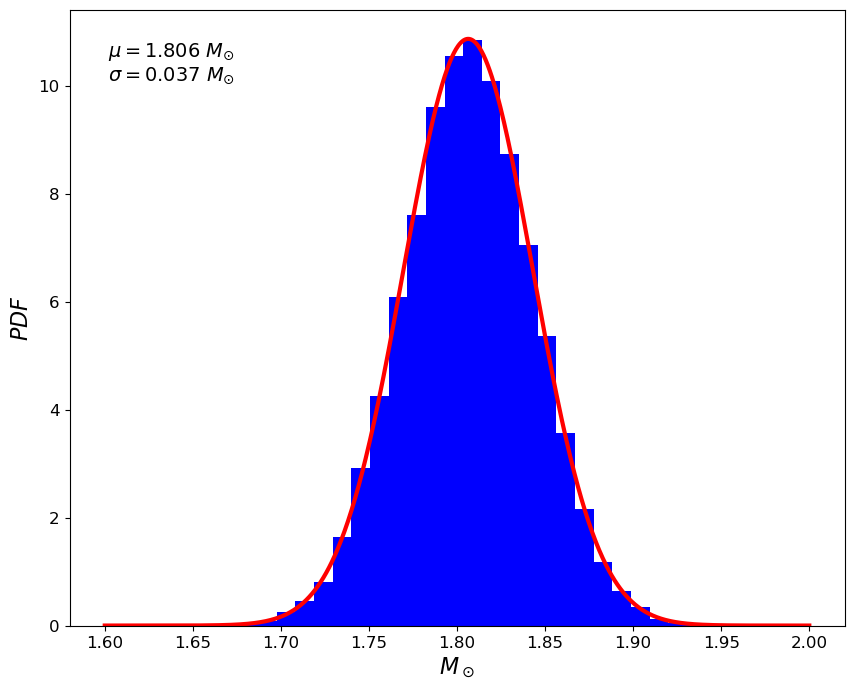}
        \caption{J0348+0432 neutron star mass distribution obtained via Monte-Carlo simulation of timing results.}
        \label{fig:NSmass}
    \end{minipage}\hfill
    \begin{minipage}{0.49\textwidth}
        \centering
        \includegraphics[width=0.75\textwidth]{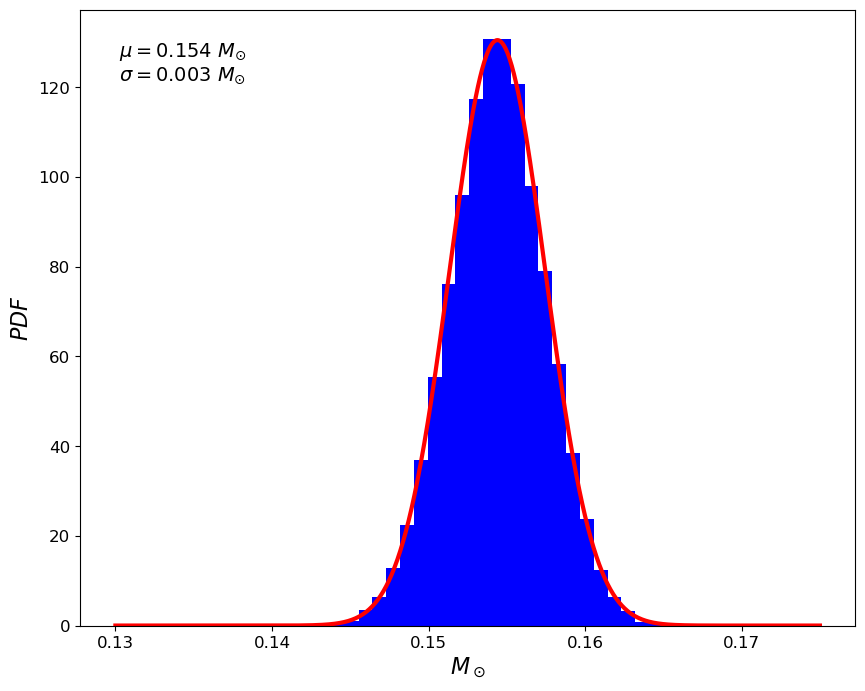}
        \caption{J0348+0432 white dwarf mass distribution obtained via Monte-Carlo simulation of timing results.}
        \label{fig:WDmass}
    \end{minipage}
\end{figure}

Our key result from this analysis is the evolution of the binary period. The period evolution ($\dot{P}_b$) of our system is measured through timing to be $-2.17(7) \times 10^{-13}$\,$[{\rm s \cdot s^{-1}}]$.
This result is $\sim$1.2$\sigma$ consistent with the value determined by \cite{Antoniadis:2013pzd}.
Additionally, the precision of our measurements is $\sim 6.4$ times greater than the original data due to a decade of added baseline.
Taking this orbital evolution and assuming it to be completely due to general relativity, we make use of the evolution equation (Eq.~\eqref{eq:Pdot}) as described in \citet{2004hpa..book.....L} and \citet{Freire:2024adf} along with the measured radial velocity measurements presented in \citet{Antoniadis:2013pzd} (which we believe to be robust, and mostly free from systematics) that give us a mass ratio of the binary (Eq.~\eqref{eq:massRatio}), we are able to determine the masses of the neutron star and its white dwarf companion:\begin{subequations}
    \begin{align}
        \dot{P}_b = -\frac{192 \pi}{5} T_{\odot}^{5/3} \left( \frac{P_b}{2 \pi}\right)^{-5/3} \left( 1 - e^2 \right)^{-7/2} \left( 1 + \frac{73}{24} e^2 + \frac{37}{96} e^4 \right) M_\mathrm{NS} M_\mathrm{WD} \left( M_\mathrm{NS} + M_\mathrm{WD} \right)^{-1/3} \,,
        \label{eq:Pdot} \\
        M_\mathrm{NS} = q \, M_\mathrm{WD} \,,
        \label{eq:massRatio} 
    \end{align}
\end{subequations}
where, $T_{\odot} \equiv \left( \mathcal{G} \mathcal{M}\right)^N_{\odot} / c^3\approx 4.925490947 \, {\rm \mu s}$ defined in~\citet{IAUInter-DivisionA-GWorkingGrouponNominalUnitsforStellarPlanetaryAstronomy:2015fjh}.
To do this, we sample each of the timing parameters with a Gaussian distribution given the mean and standard deviation as presented in Table~\ref{tab:TimingResults}.  
We then perform a Monte-Carlo simulation to give us a distribution of masses.
The results of the sampling scheme are presented in Figs.~\ref{fig:NSmass} \&~\ref{fig:WDmass}.
We find that the new mass measurements give a neutron star mass $M_\mathrm{NS} = 1.806(37)$\,${\rm M_\odot}$ and a white dwarf mass $M_\mathrm{WD} = 0.154(3)$\,${\rm M_\odot}$.
The values of $T_{eff}$ and $\log g$ observed for the white-dwarf companion are representative of those in recent surveys for extremely low-mass WDs~\citep{2023ApJ...950..141K,2022ApJ...933...94B} and recent cooling models~\citep{2016A&A...595A..35I} imply that the mass is likely considerably lower than the upper range derived in~\citet{Antoniadis:2013pzd}.

These masses are significantly different from those of the~\citet{Antoniadis:2013pzd} paper.
If we make use of Eq.~\eqref{eq:Pdot} with the values for $M_{\rm NS}$, $M_{\rm WD}$, $P_b$, and $e$ presented in the former work we find that the calculated $\dot{P}_{\rm B,xs}=-2.593(78)\times 10^{-13}$\,$[{\rm s \cdot s^{-1}}]$.
 From here on, $\dot{P}_{\rm B,xs}$ will be a reference to the analytical orbital evolution rate assuming the physical parameters from the previous research.
Comparing this value with our own value of $\dot{P}_b$ and adding the uncertainties in quadrature shows that there is a $\sim 4 \sigma$ discrepancy between the two values.

We may also investigate two kinematic effects which could lead to changes in the observed $\dot{P}_b$.
The first is the ``Shklovskii Effect,"~\citep{1970SvA....13..562S} which can be thought of as an artificial period derivative due to transverse motion of the pulsar at some distance away from Earth.
A separate effect is due to the difference in Galactic acceleration of the solar system barycenter (SSB) and the center of mass of the pulsar system.~\citep{1991ApJ...366..501D,1995ApJ...441..429N,2009MNRAS.400..805L}.
Knowing the galactic location of the pulsar, distance between the SSB and pulsar, and a model of Galactic potential we may be able to calculate a change in the orbital period by calculating the relative accelerations, which leads to a time-varying doppler shift~\citep{1992RSPTA.341...39P}.
We make use of the \textsc{presto}\footnote{See~\citep{2011ascl.soft07017R} or \href{https://github.com/scottransom/presto}{https://github.com/scottransom/presto}} software package, which contains the necessary functions to determine these orbital evolution found in~\citet{1970SvA....13..562S,1995ApJ...441..429N} to solve the equations for the Shklovskii Effect ($\dot{P}^{\rm Gal}_b$), as well as the Galactic acceleration effects both parallel (${}_\parallel\dot{P}_b$) and perpendicular\footnote{We use the Galactic potential model of~\citet{1989MNRAS.239..571K,1989MNRAS.239..605K} with a local mass density of $\rho_{\odot}=10^{-2} \, {\rm M_\odot \, pc}^2$ and a total disk column density of $\Sigma=46 \, {\rm M_\odot \, pc}^{-2}$ as was done in~\citet{1995ApJ...441..429N}.} (${}_\perp \dot{P}_b^{\rm Gal}$) to the galactic plane:
\begin{subequations}
    \begin{align}
        \frac{\dot{P}^{\rm Shk}_b}{P_b} 
        &= \frac{\mu^2 d}{c} \,, \\
        \frac{{}_\parallel\dot{P}^{\rm Gal}_b}{P_b} &= -\frac{V_o^2}{c R_o} \cos{b} \left[ \cos{l} + \frac{\left(d/R_o\right) \cos{b} - \cos{l}}{1+ \left( d/R_o\right)^2 \cos^2{b} -2 \left( d/R_o \right) \cos{b} \cos{l}}\right] \,, \\
        \frac{{}_{\perp}\dot{P}^{\rm Gal}_b}{P_b} &=1.08 \times 10^{-19} \left[ \frac{1.25 }{\sqrt{d^2 \sin^2{b} + 0.0324}}+0.58\right] \left( d \sin^2{b}\right) \, [{\rm s^{-1}}] \,,
    \end{align}
\end{subequations}
where $\mu$ is the total proper motion of the pulsar, $d=2.1(1)$ kpc~\citep{Antoniadis:2013pzd} is the distance from the Earth to the pulsar, $V_o \approx 240$ km/s~\citep{2014ApJ...783..130R} is the orbital speed of the Sun around the center of the galaxy, $R_o=8.34$ kpc~\citep{2014ApJ...783..130R}, and $b$ and $l$ are the galactic longitude and latitude of the pulsar.
Using the location information found in Tab.~\ref{tab:TimingResults}, the orbital effects are found to be ${\dot{P}_b^{\rm Shk}}/{P_b} \approx 9.3(6) \times 10^{-20} \, [{\rm s}^{-1}]$ ,${{}_\parallel\dot{P}_b^{\rm Gal}}/{P_b} \approx 9.9(3) \times 10^{-20} \, [{\rm s}^{-1}]$, and ${}_\perp \dot{P}_b^{\rm Gal} \approx -2.47(5) \times 10^{-19} \, [{\rm s}^{-1}]$,  all of which are several orders of magnitude smaller than our value of ${\dot{P}_b}/{P_b} \approx - 2 \times 10^{-17} \, [{\rm s}^{-1}]$. 
While the above galactic acceleration equations are highly simplified, they shouldn't be off by more than a factor of a few from the real value, and what is important is simply that they have a very small magnitude.
We note that the magnitude of the proper motion terms used in the calculations above agree with the results published in the Gaia DR3 release~\citep{2016A&A...595A...1G,2023A&A...674A...1G}\footnote{J0348+0432 is identified as Gaia DR3 3273288485744249344.}, though the Gaia experiment itself only constrained the parallax to a distance of $-0.035 \pm 0.784 \,{\rm mas}$~\citep{2024PhRvD.109l3015M}.
These results indicate that there is no significant contamination in our data from the kinematic effects we investigated.

Calculating higher derivatives of a couple of the physical parameters may also prove advantageous in determining the validity of our observations.
We find that the second derivative of the pulsar spin frequency is $1.4(1.0) \times 10^{-28}$\,$[{\rm Hz \cdot s^{-2}}]$.  This value is minuscule, and combined with the flat timing residuals, implies that timing noise is highly unlikely to contaminate our results.
The derivative of the projected semi-major axis of the binary orbit was also fitted to the data.
This could let us know if any extra effects, such as projection effects from proper motion, were present which we may have missed that could influence $\dot{P}_b$.
From fitting the data, we determined this value to be $-2.5(3.3) \times 10^{-15}$\,$[{\rm ls \cdot s^{-1}}]$, a non-detection, and too small to affect our value of $\dot{P}_b$ in any meaningful way.

\section{Discussion}
The results of our updated timing analysis lead to a significant difference in the inferred pulsar mass originally determined by \citet{Antoniadis:2013pzd}. The updated estimate of $M_\mathrm{NS} = 1.806(37)\,\textrm{M}_\odot$, which differs from the original estimate of $2.01(4)\,\textrm{ M}_\odot$ by at least $\sim 5 \sigma$, challenges the notion that PSR J0348+4232 is one of the highest-mass neutron stars known. Given the significance of this change, we address different aspects of our result in relation to prior analysis methods.

\subsection{Differences Between Mass-Estimation Methods}
\cite{Antoniadis:2013pzd} presented a radio timing solution that exhibited unambiguous orbital decay of the system; they argued that the observed $\dot{P}_b$ must be due to gravitational-wave emission after carefully considering all other known forms of intrinsic or apparent variations in $P_b$.
We confirmed their estimate of orbital decay with our combined Arecibo and CHIME/Pulsar data sets, finding that our latest estimate differs only by $\sim$1.2$\sigma$ from the value previously determined.
We consider this consistency to be a confirmation of the system behavior as gauged from radio timing, given the extension of the data set by at least 10 years. However, the radio-timing estimate of $\dot{P}_b = -2.73(45)\times10^{-13}$\,$[{\rm s \cdot s^{-1}}]$~\cite{Antoniadis:2013pzd} possessed weaker statistical significance than the physical parameters derived from modeling of optical spectra of the white-dwarf companion. Therefore, the observed $\dot{P}_b$ was not used by \cite{Antoniadis:2013pzd} for their constraint on the mass of PSR J0348+0432.

\cite{Antoniadis:2013pzd} instead relied on high-precision measurements of brightness and spectral parameters of the white dwarf to derive their initial estimate of $M_\mathrm{NS}$. The procedure consisted of: measuring surface gravity -- $\log{g}$, where $g=GM_\mathrm{WD}/R_\mathrm{WD}^2$ -- through direct fitting of lines observed in optical spectra, which requires an assumption of surface-envelope composition; assuming a finite-temperature relationship between $M_\mathrm{WD}$ and $R_\mathrm{WD}$, arising from the white-dwarf equations of state, to derive $M_\mathrm{WD}$; and using the Keplerian mass function to derive $M_\mathrm{NS}$. Past work has shown such a method to yield self-consistent properties of the white dwarf when, e.g., deriving $R_\mathrm{WD}$ instead from integrated-flux data through the distance modulus, which depends on {\it a priori} knowledge of the distance and optical reddening \citep{Bassa:2006b,Antoniadis:2012}.

Either method nonetheless invokes several assumptions that can introduce systematic uncertainties into any estimation of $M_\mathrm{NS}.$ It is therefore likely that one or a combination of these assumptions lead to a composite systematic uncertainty whose magnitude cannot not easily assessed. By contrast, the use of $q$ and $\dot{P}_b$ in determining the masses only requires the assumption that general relativity explain the observed orbital decay, which is well-tested at our current level of precision \cite[e.g.,][]{Stairs:2003} for NS$-$WD binaries. This combination was also performed by \cite{Antoniadis:2013pzd} and yielded a less-stringent estimate of $M_\mathrm{NS}$ that is nonetheless consistent with our improved result. The consistency in radio-timing models further points to the notion that the optical-only derivation of $M_\mathrm{WD}$ (which when used in combination with the measured velocity ratio can yield $M_\mathrm{NS}$) suffers from a systematic bias, and we therefore consider the $q-\dot{P}_b$ estimate of $M_\mathrm{NS}$ in our analysis to be more robust.

\subsection{Monte-Carlo Estimation of Mass Uncertainties}
We also investigated what was the largest source of error with our simulations.
To do this, we run two simulations where one assumes a perfect $\dot{P}_b$ with the mass ratio $q$ taken from~\citet{Antoniadis:2013pzd}, and the other assumes a perfect measurement on the mass ratio, while we keep the $\dot{P}_b$ we determine via our observations.
By comparing these two scenarios, we were able to deduce that the uncertainty on the evolution of the binary period ($\dot{P}_b$) was the largest source of inaccuracy in our system.
The expanded timing baseline that we include in this work which gives over ten years of timing with the included CHIME/Pulsar data helps in reducing the errors and providing us with a more accurate estimate of $\dot{P}_b$.
This should be the case given that the error on $\dot{P}_b$ scales as $T^{5/2}$ where $T$ is the observation time~\citep{2004hpa..book.....L}.
We expect that continued study of the system J0348+0432 will further hone the value of $\dot{P}_b$ and thus provide more accurate mass measurements.
A cursory calculation shows that approximately 23 more years of continued observations of J0348+0432 with CHIME/Pulsar should reduce the error on the $\dot{P}_b$ measurement to the point where the dominant source of uncertainty in the system is the optical measurements used for the determination of the mass ratio.

\subsection{Mass Loss as a Cause of Orbital Decay and DM Variations}
\label{sec:mLossandDM}

\begin{figure}
    \centering
    \includegraphics[width=\textwidth]{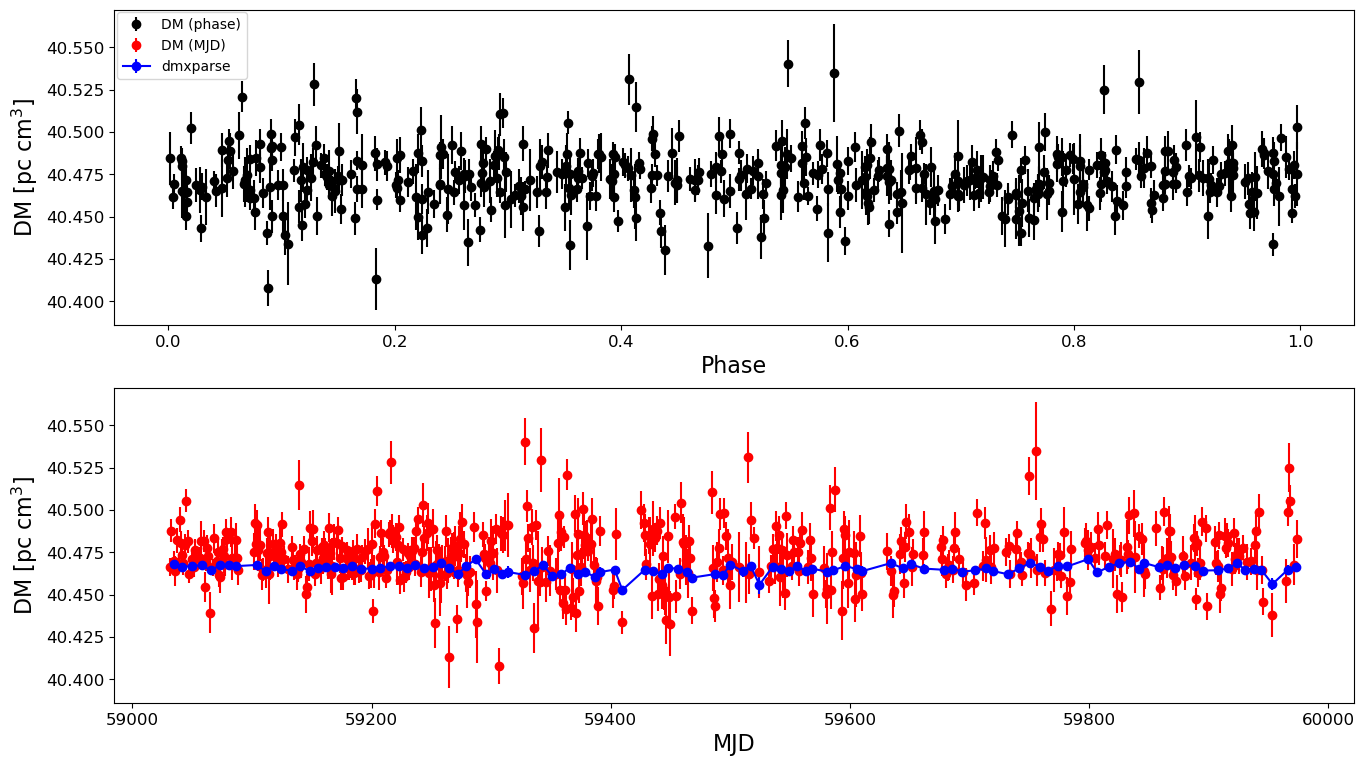}
    \caption{Top: The dispersion measure against the orbital phase (black) for J0348+0432 using the new CHIME/Pulsar data.  The CHIME data was frequency scrunched to 8 channels before applying the {\tt PulsePortraiture} procedure.  As seen in the plot, there is no significant trend with phase which could indicate the interaction between the orbital bodies and any extraneous matter which could affect the binary parameters.  Note that superior conjunction, where the companion is in front of the pulsar, and where we would most likely see excess dispersion, occurs at orbital phase of 0.25.  Bottom:  Same as top, except now compared with time (red).   We have also included the results from the DMX analysis (blue), which agree with the {\tt PulsePortraiture} procedure.  No trends are noticed to indicate any mass loss leading to an improper measure of the orbital parameters.}
    \label{fig:DM_Evolution}
\end{figure}

It is possible that the measured $\dot{P}_b$ is not solely from gravitational effects, but has a contribution due to some other effect, like mass loss from the system~\citep{1924MNRAS..85....2J,1928asco.book.....J}. As the binary system sheds mass -- either from the pulsar or its companion -- there is an expectation that this will be in the form of relativistic particles and electromagnetic fields~\citep{1976ApJ...207..574S}.
As such, this should leave an imprint on the dispersion measure (DM) calculated for the system as the pulsed signal will traverse and interact with the local plasma. Such orbital DM variations have been observed with CHIME data in the PSR J1641+8049 ``black-widow" binary system, where high-order variations in the orbital parameters due to mass loss are also measured \citep{kzk+24}.
Noting how the dispersion measure changes over time should provide evidence that there is an additional effect of mass loss present, which in theory could account for a variation in the $\dot{P}_b$ observed.
\citet{Antoniadis:2013pzd} also discussed the possibility of mass loss contributing to the change in orbital period.
Their original results placed an upper limit of $\dot{P}_b^{\dot{m}}<0.4 \times 10^{-16}$, roughly 500 times less than the orbital evolution rate governed by gravitational effects.
We made use of the {\tt PulsePortraiture}\footnote{\href{https://github.com/pennucci/PulsePortraiture}{https://github.com/pennucci/PulsePortraiture}} software package to track the evolution of the DMs using wideband timing~\citep{Pennucci:2014dja,Pennucci:2018zow}.
The DM was calculated as explained in~\ref{subsec:DM_CHIME}, where the original CHIME data was frequency scrunched to 8 channels.
Figure~\ref{fig:DM_Evolution} shows the results of our analysis.
We note that within the CHIME/Pulsar data, there is no DM trend which would indicate the presence of particles which would result from the mass loss of the binary system.

As an additional check, we used {\tt PINT} to derive DMs a different way.
Taking advantage of the CHIME data which was frequency scrunched to 8 channels that we used in the {\tt PulsePortraiture} analysis, we use DMX\footnote{See~\citet{NANOGrav:2023hde} Sec. 4 for an explanation of the DMX method.} values which we derive via {\tt dmxparse}, a utility originally used in {\tt tempo} but updated with the newer software to ensure each DMX bin contained at least one TOA.
Figure~\ref{fig:DM_Evolution} shows the results of the DMX procedure in the bottom portion of the plot.
Here, there are no trends or significant deviations from the mean DM value for either procedure discussed.
This suggests that the presence of significant additional ionized gas near the binary system is unlikely.
Following the work and procedures laid out in~\citet{2000ApJ...541..335K,Ransom:2003qv,Possenti:2003nr}, we can substantiate this claim by making a rough estimate of the expected mass loss given an upper limit a $\Delta{\rm DM}$ that we do not see.
Via Fig.~\ref{fig:DM_Evolution}, we fail to see an excess DM around conjunction (orbital phase = 0.25) of more than about $0.03 \, {\rm pc \cdot cm^{3}}$.
If we assume that the WD is emitting a fully ionized plasma isotropically, we may calculate the mass loss via $\dot{M}=4 \pi R_{\rm E}^2 \rho_{\rm E} v_w$, where $R_{\rm E}$ is the eclipsing region of the binary, $\rho_{\rm E}$ is the plasma density, and $v_w$ is the escape velocity for the WD.
We assume our values to be $R_{\rm E} \approx 0.3 R_{\odot}$, $\rho_{\rm E} \approx 3 \times 10^{-18}\, [{\rm g \cdot cm^{-3}}]$ where our excess DM was integrated over a column the length of the semi-major axis of the orbit, and $v_w \approx 10^{8}\,[{\rm cm \cdot s}]$.
This gives us $\dot{M} \approx 3 \times 10^{-14} \, [{\rm M_\odot / yr}]$ to be the estimated mass loss of the companion WD.

It is still useful to check and see what the expected mass loss would be if $\dot{P}_{\rm B,xs}$ is the correct orbital evolution rate given the original masses from~\citet{Antoniadis:2013pzd}.
The change in orbital period assuming a system is losing mass can be given by~\citep{1991ApJ...366..501D}
\begin{equation}
    \label{eq:OrbitalLossMass}
    \dot{P}^{\dot{m}}_{b}=2 P_b \left( \frac{\dot{M}_{\rm NS}+\dot{M}_{\rm WD}}{M_{\rm NS} + M_{\rm WD}} \right)\,.
\end{equation}
Given our lower value for $\dot{P}_b$, we must obtain an evolution due to mass loss of $\dot{P}^{\dot{m}}_b=-4.2(10) \times 10^{-14}$ in order to agree with $\dot{P}_{\rm B,xs}$.
Taking the values from~\citet{Antoniadis:2013pzd} for the masses, period, as well as $\dot{M}_{\rm NS}$ which they find to be $2.82(5) \times 10^{-15} \, {\rm M_\odot} {\rm yr}^{-1}$, we can find the mass loss of the white dwarf must be $\dot{M}_{\rm WD} \approx  -1.66(40) \times 10^{-10} \, {\rm M_\odot} \, {\rm yr}^{-1}$.
This value is extremely large compared with already observed mass loss rates for black widow and redback systems~\citep{1990ApJ...351..642F} as well as our own estimate for the mass loss calculated above.
We may therefore safely assume that the $\dot{P}_b$ we have found in this work is not dominated by the mass loss of the system.

\begin{figure}
    \centering
    \includegraphics[width=0.65\textwidth]{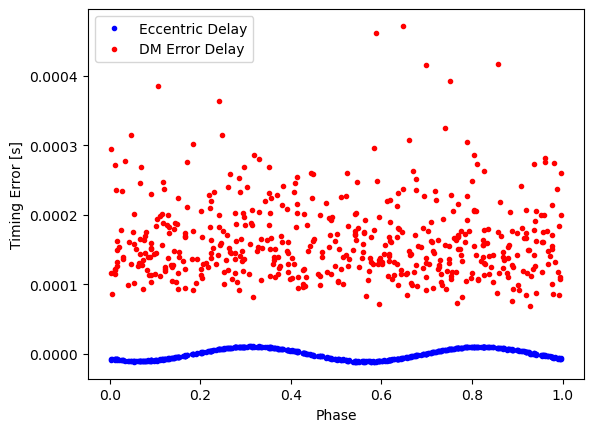}
    \caption{A comparison of the delays calculated between the DM and the assumption that the orbit is eccentric.  We note that the DM error is roughly an order of magnitude larger than the delays induced by the eccentricity of the orbit.  This is a further evidence that there is no extraneous DM variations leading to a misunderstanding of our system.}
    \label{fig:Error_Eval}
\end{figure}

We also investigated if more periodic DM variations from a smoother wind could cause a mis-estimation of the eccentricity of the binary, and mask potential mass loss.  The eccentricity of the binary and an ionized wind should cause timing delays as a function of the orbital period.  If delays from a wind are dominated by the DM errors, they could in theory skew the orbital parameters of the binary which could induce corrections to the $\dot{P}_b$ and the NS mass.  We find the delay caused by the eccentricity of the orbit by taking the difference between the orbital delays of the J0348+0432 system assuming an eccentric and circular orbit.  
For this analysis, we again made use of {\tt PulsePortraiture} and the frequency scrunched data which is described in~\ref{subsec:DM_CHIME}.
This procedure allow us to record a DM and DM timing error for each TOA.
The results of this analysis is shown in Fig.~\ref{fig:Error_Eval}.
We find that it is unlikely that systematic DM errors as a function of orbital phase would cause inaccuracies in eccentricity because the DM errors are roughly an order of magnitude larger.  Therefore, we may conclude that our assumption that the orbit is eccentric is correct and the calculated orbital parameters are valid.  This includes our new values of the $\dot{P}_b$ and $M_{\rm NS}$.

\section{Conclusions}
The work presented here establishes a new mass measurement for the pulsar J0348+0432.
Using a much longer timing baseline with the inclusion of CHIME/Pulsar data allows for a higher precision measurement of the rate of orbital evolution.
Together with the original spectroscopic measurements and inference of the binary mass ratio, we deduce that the $M_{\rm NS}=1.806(37)$\,${\rm M_\odot}$.
We believe this result to be robust as it relies solely on observed quantities rather than a model of broadband optical emission.
Tests were performed on DM variations to see if any mass shedding is present which can influence our measurement of $\dot{P}_b$.
From these, we determine that our DM is stable, indicating that the system is most likely not shedding significant mass which could affect the orbital evolution.
We looked into various kinematic effects which could be present to cause a deviation in the measured $\dot{P}_b$.
We investigated both the Shklovskii Effect and the effect of galactic acceleration and found that neither is a significant contributor to the overall binary evolution of the system.
We also investigated whether a mis-estimation of the eccentricity of the system could have masked ionized mass loss from the system, thereby contaminating our measurement of $\dot{P}_b$.
We find that our errors on the value of the eccentricity are much smaller than our uncertainties from our DM measurements, and therefore will not have any significant influence on our $\dot{P}_b$.
Given that we have shown the robustness of our $\dot{P}_b$ measurements, we conclude that our updated mass measurements are valid.
With this in mind, J0348+0432 promises to be an exciting object for future study, namely as a test-bed for modified gravity.
Such work is already underway (Paulo Freire, private communication), which will include additional timing data from Effelsberg and Arecibo.
It is well known that modified theories of gravity may induce alternative modes of gravitational waves~\citep{1975ApJ...196L..59E,Will:1993hxu,Berti:2015itd}.
These additional modes carry energy with them, causing the binary to inspiral at a different rate than that induced by general relativity alone.
This change in $\dot{P}_b$ is a measurable effect which can be used to place bounds on certain alternative theories of gravity.
We expect to make use of follow-up observations of this system in order to place bounds on the strength of the alternative gravitational wave modes by measuring the $\dot{P}_b$ and comparing the results with values expected from general relativity.

\begin{table}[]
    \centering
    \begin{tabular}{|c|c|c|c|}
    \hline
    \multicolumn{4}{|c|}{Timing Parameters} \\
    \hline
         Name & Our Values & Antoniadis \textit{et. al.} & CHIME Only\\
         \hline
         Right Ascension (J2000) & 03:48:43.638483(11)& 
         03:48:43.639000(4) & 03:48:43.64173(25)\\
         Declination (J2000) & $04^{\circ} 32' 11''.45456(49)$& $04^{\circ} 32' 11''.4580(2)$ & $04^{\circ} 32' 11''.4626(96)$\\
         Proper Motion Right Ascension $[\rm{mas} \cdot \rm{yr}^{-1}]$ & 4.12(7)& 4.04(16) & 3.0(32)\\
         Proper Motion Declination $[\rm{mas} \cdot \rm{yr}^{-1}]$ & 1.12(24)& 3.5(6) & -6(8)\\
         Pulsar Frequency [\rm Hz]& 25.5606365999036(9)& 25.5606361937675(4) & 25.560636547705(12)\\
         First Derivative of Pulsar Frequency $[\rm{Hz} \cdot \rm{s}^{-1}]$ & $-1.57323(18)\times 10^{-16}$& $-1.5729(3) \times 10^{-16}$ & $-1.5696(35) \times 10^{-16}$\\
         Dispersion Measure $[\rm{cm}^{-3} \cdot \rm{pc}]$& 40.46329(9)& 40.46313(11) & -\footnote{As our CHIME Only model was integrated over time and frequency, an accurate determination of the DM was not obtained.}\\
         Binary Model & ELL1 & ELL1 & ELL1 \\
         Orbital Period ($P_b$) [\rm{day}] & 0.102424061371(9)& 0.102424062722(7) & 0.1024240608(5)\\
         Orbital Evolution Rate ($\dot{P}_b$)& $-2.17(7) \times 10^{-13}$& $-2.73(45) \times 10^{-13}$ & $-5(12) \times 10^{-13}$\\
         Projected Semi-Major Axis [ls] & 0.14098106(33)& 0.14097938(7) & 0.140982(5)\\
         Time of Ascending Node [MJD] & 54889.70532365(6)& 56000.084771047(11) & 54889.70513(15)\\
         $\eta = e \sin(\omega)$ & $4.6(1) \times 10^{-6}$& $1.9(10) \times 10^{-6}$ & 0.00010(5)\\
         $\kappa = e \cos(\omega)$ & $0.3(10) \times 10^{-6}$& $1.4(10) \times 10^{-6}$ & $2(5) \times 10^{-5}$\\
         Mass Ratio ($q={\rm M_{\mathrm{NS}}/M_{\mathrm{WD}}}$) & - & 11.70(13) & - \\
         \hline
         \multicolumn{4}{|c|}{Derived Parameters} \\
         \hline
         Spin Period [ms] & 39.1226562801558(13)& 39.1226569017806(5) & 39.122656360050(18)\\
         First Derivative of Spin Period $[{ \rm s \cdot s^{-1}}]$& 2.40795(28) $\times 10^{-19}$& 2.4073(4) $\times 10^{-19}$ & 2.402(5) $\times 10^{-19}$\\
         Characteristic Age [Gyr] & 2.574& 2.6 & 2.58\\
        Eccentricity (${\rm e}$) & $4.7(10)\times 10^{-6}$& - & 0.00010(5)\\
         Transverse Magnetic Field at Poles [G] & 3.11 $\times 10^{9}$& $\sim 2 \times 10^{9}$ & $3.1 \times 10^{9}$\\
         Mass Function $[{\rm M_\odot}]$ & 0.0002867881(20)& 0.000286778(4) & 0.000286793(21)\\
         \hline
    \end{tabular}
    \caption{All timing and derived parameters from our analysis.  We include the total combined values in the first column, the archival values from~\citet{Antoniadis:2013pzd} in the second column, and the analysis of only the newly added CHIME/Pulsar data in the third column.}
    \label{tab:TimingResults}
\end{table}

\section*{Acknowledgements}

We acknowledge that CHIME is located on the traditional, ancestral, and unceded territory of the Syilx/Okanagan people. We are grateful to the staff of the Dominion Radio Astrophysical Observatory, which is operated by the National Research Council of Canada. CHIME operations are funded by a grant from the NSERC Alliance Program and by support from McGill University, University of British Columbia, and University of Toronto. CHIME was funded by a grant from the Canada Foundation for Innovation (CFI) 2012 Leading Edge Fund (Project 31170) and by contributions from the provinces of British Columbia, Québec and Ontario. The CHIME/FRB Project, which enabled development in common with the CHIME/Pulsar instrument, was funded by a grant from the CFI 2015 Innovation Fund (Project 33213) and by contributions from the provinces of British Columbia and Québec, and by the Dunlap Institute for Astronomy and Astrophysics at the University of Toronto. Additional support was provided by the Canadian Institute for Advanced Research (CIFAR), the Trottier Space Institute at McGill University, and the University of British Columbia.  The CHIME/Pulsar instrument hardware is funded by the Natural Sciences and Engineering Research Council (NSERC) Research Tools and Instruments (RTI-1) grant EQPEQ 458893-2014.

We would like to sincerely thank Paulo Freire for his discussions and inputs with the initial drafts of the manuscript.

Computational resources were provided by the Link HPC cluster and cyber-infrastructure, which is maintained by the Center for Gravitational Waves and Cosmology at West Virginia University and funded in part by NFS IIA-1458952 and NFS PHY-2020265.  Additional computing and data storage resources were provided by the Digital Research Alliance of Canada.
The National Radio Astronomy Observatory is a facility of the National Science Foundation operated under cooperative agreement by Associated Universities, Inc.
The Arecibo Observatory wss operated by SRI International under a cooperative agreement with the National Science Foundation (AST-1100968), and in alliance with Ana G. Méndez-Universidad Metropolitana, and the Universities Space Research Association.  

SMR is a CIFAR Fellow and he and AS are supported by the NSF Physics Frontiers Center award 2020265.
Pulsar research at UBC is supported by an NSERC Discover Grant and by the Canadian Institute for Advanced Research.
D.C.G is supported by the National Science Foundation under Grant No. 2406919.
K.W.M. holds the Adam J. Burgasser Chair in Astrophysics.

\bibliography{bibliography}{}

\begin{thebibliography}{}
\expandafter\ifx\csname natexlab\endcsname\relax\def\natexlab#1{#1}\fi
\providecommand{\url}[1]{\href{#1}{#1}}
\providecommand{\dodoi}[1]{doi:~\href{http://doi.org/#1}{\nolinkurl{#1}}}
\providecommand{\doeprint}[1]{\href{http://ascl.net/#1}{\nolinkurl{http://ascl.net/#1}}}
\providecommand{\doarXiv}[1]{\href{https://arxiv.org/abs/#1}{\nolinkurl{https://arxiv.org/abs/#1}}}

\bibitem[{Agazie {et~al.}(2023)}]{NANOGrav:2023hde}
Agazie, G., {et~al.} 2023, Astrophys. J. Lett., 951, L9, \dodoi{10.3847/2041-8213/acda9a}

\bibitem[{Amiri {et~al.}(2021)Amiri, Bandura, Boyle, Brar, Cliche, Crowter, Cubranic, Demorest, Denman, Dobbs, Dong, Fandino, Fonseca, Good, Halpern, Hill, Höfer, Kaspi, Landecker, Leung, Lin, Luo, Masui, McKee, Mena-Parra, Meyers, Michilli, Naidu, Newburgh, Ng, Patel, Pinsonneault-Marotte, Ransom, Renard, Scholz, Shaw, Sikora, Stairs, Tan, Tendulkar, Tretyakov, Vanderlinde, Wang, \& Wang}]{Amiri_2021}
Amiri, M., Bandura, K.~M., Boyle, P.~J., {et~al.} 2021, The Astrophysical Journal Supplement Series, 255, 5, \dodoi{10.3847/1538-4365/abfdcb}

\bibitem[{Amiri {et~al.}(2022)}]{CHIME:2022dwe}
Amiri, M., {et~al.} 2022, Astrophys. J. Supp., 261, 29, \dodoi{10.3847/1538-4365/ac6fd9}

\bibitem[{{Antoniadis} {et~al.}(2012){Antoniadis}, {van Kerkwijk}, {Koester}, {Freire}, {Wex}, {Tauris}, {Kramer}, \& {Bassa}}]{Antoniadis:2012}
{Antoniadis}, J., {van Kerkwijk}, M.~H., {Koester}, D., {et~al.} 2012, \mnras, 423, 3316, \dodoi{10.1111/j.1365-2966.2012.21124.x}

\bibitem[{Antoniadis {et~al.}(2013)}]{Antoniadis:2013pzd}
Antoniadis, J., {et~al.} 2013, Science, 340, 6131, \dodoi{10.1126/science.1233232}

\bibitem[{{Bassa} {et~al.}(2006){Bassa}, {van Kerkwijk}, {Koester}, \& {Verbunt}}]{Bassa:2006b}
{Bassa}, C.~G., {van Kerkwijk}, M.~H., {Koester}, D., \& {Verbunt}, F. 2006, \aap, 456, 295, \dodoi{10.1051/0004-6361:20065181}

\bibitem[{Berti {et~al.}(2015)}]{Berti:2015itd}
Berti, E., {et~al.} 2015, Class. Quant. Grav., 32, 243001, \dodoi{10.1088/0264-9381/32/24/243001}

\bibitem[{{Brown} {et~al.}(2022){Brown}, {Kilic}, {Kosakowski}, \& {Gianninas}}]{2022ApJ...933...94B}
{Brown}, W.~R., {Kilic}, M., {Kosakowski}, A., \& {Gianninas}, A. 2022, \apj, 933, 94, \dodoi{10.3847/1538-4357/ac72ac}

\bibitem[{Chamel {et~al.}(2013)Chamel, Haensel, Zdunik, \& Fantina}]{Chamel:2013efa}
Chamel, N., Haensel, P., Zdunik, J.~L., \& Fantina, A.~F. 2013, Int. J. Mod. Phys. E, 22, 1330018, \dodoi{10.1142/S021830131330018X}

\bibitem[{{CHIME/Pulsar Collaboration} {et~al.}(2021){CHIME/Pulsar Collaboration}, {Amiri}, {Bandura}, {Boyle}, {Brar}, {Cliche}, {Crowter}, {Cubranic}, {Demorest}, {Denman}, {Dobbs}, {Dong}, {Fandino}, {Fonseca}, {Good}, {Halpern}, {Hill}, {H{\"o}fer}, {Kaspi}, {Landecker}, {Leung}, {Lin}, {Luo}, {Masui}, {McKee}, {Mena-Parra}, {Meyers}, {Michilli}, {Naidu}, {Newburgh}, {Ng}, {Patel}, {Pinsonneault-Marotte}, {Ransom}, {Renard}, {Scholz}, {Shaw}, {Sikora}, {Stairs}, {Tan}, {Tendulkar}, {Tretyakov}, {Vanderlinde}, {Wang}, \& {Wang}}]{chimepulsar21}
{CHIME/Pulsar Collaboration}, {Amiri}, M., {Bandura}, K.~M., {et~al.} 2021, ApJS, 255, 5, \dodoi{10.3847/1538-4365/abfdcb}

\bibitem[{Cromartie {et~al.}(2019)}]{NANOGrav:2019jur}
Cromartie, H.~T., {et~al.} 2019, Nature Astron., 4, 72, \dodoi{10.1038/s41550-019-0880-2}

\bibitem[{{Damour} \& {Taylor}(1991)}]{1991ApJ...366..501D}
{Damour}, T., \& {Taylor}, J.~H. 1991, \apj, 366, 501, \dodoi{10.1086/169585}

\bibitem[{Demorest {et~al.}(2010)Demorest, Pennucci, Ransom, Roberts, \& Hessels}]{Demorest:2010bx}
Demorest, P., Pennucci, T., Ransom, S., Roberts, M., \& Hessels, J. 2010, Nature, 467, 1081, \dodoi{10.1038/nature09466}

\bibitem[{{Eardley}(1975)}]{1975ApJ...196L..59E}
{Eardley}, D.~M. 1975, \apjl, 196, L59, \dodoi{10.1086/181744}

\bibitem[{Fonseca {et~al.}(2021)}]{Fonseca:2021wxt}
Fonseca, E., {et~al.} 2021, Astrophys. J. Lett., 915, L12, \dodoi{10.3847/2041-8213/ac03b8}

\bibitem[{Freire \& Wex(2024)}]{Freire:2024adf}
Freire, P. C.~C., \& Wex, N. 2024, Living Rev. Rel., 27, 5, \dodoi{10.1007/s41114-024-00051-y}

\bibitem[{{Fruchter} {et~al.}(1990){Fruchter}, {Berman}, {Bower}, {Convery}, {Goss}, {Hankins}, {Klein}, {Nice}, {Ryba}, {Stinebring}, {Taylor}, {Thorsett}, \& {Weisberg}}]{1990ApJ...351..642F}
{Fruchter}, A.~S., {Berman}, G., {Bower}, G., {et~al.} 1990, \apj, 351, 642, \dodoi{10.1086/168502}

\bibitem[{{Gaia Collaboration} {et~al.}(2016){Gaia Collaboration}, {Prusti}, {de Bruijne}, {Brown}, {Vallenari}, {Babusiaux}, {Bailer-Jones}, {Bastian}, {Biermann}, {Evans}, {Eyer}, {Jansen}, {Jordi}, {Klioner}, {Lammers}, {Lindegren}, {Luri}, {Mignard}, {Milligan}, {Panem}, {Poinsignon}, {Pourbaix}, {Randich}, {Sarri}, {Sartoretti}, {Siddiqui}, {Soubiran}, {Valette}, {van Leeuwen}, {Walton}, {Aerts}, {Arenou}, {Cropper}, {Drimmel}, {H{\o}g}, {Katz}, {Lattanzi}, {O'Mullane}, {Grebel}, {Holland}, {Huc}, {Passot}, {Bramante}, {Cacciari}, {Casta{\~n}eda}, {Chaoul}, {Cheek}, {De Angeli}, {Fabricius}, {Guerra}, {Hern{\'a}ndez}, {Jean-Antoine-Piccolo}, {Masana}, {Messineo}, {Mowlavi}, {Nienartowicz}, {Ord{\'o}{\~n}ez-Blanco}, {Panuzzo}, {Portell}, {Richards}, {Riello}, {Seabroke}, {Tanga}, {Th{\'e}venin}, {Torra}, {Els}, {Gracia-Abril}, {Comoretto}, {Garcia-Reinaldos}, {Lock}, {Mercier}, {Altmann}, {Andrae}, {Astraatmadja}, {Bellas-Velidis}, {Benson}, {Berthier}, {Blomme}, {Busso}, {Carry}, {Cellino}, {Clementini},
  {Cowell}, {Creevey}, {Cuypers}, {Davidson}, {De Ridder}, {de Torres}, {Delchambre}, {Dell'Oro}, {Ducourant}, {Fr{\'e}mat}, {Garc{\'\i}a-Torres}, {Gosset}, {Halbwachs}, {Hambly}, {Harrison}, {Hauser}, {Hestroffer}, {Hodgkin}, {Huckle}, {Hutton}, {Jasniewicz}, {Jordan}, {Kontizas}, {Korn}, {Lanzafame}, {Manteiga}, {Moitinho}, {Muinonen}, {Osinde}, {Pancino}, {Pauwels}, {Petit}, {Recio-Blanco}, {Robin}, {Sarro}, {Siopis}, {Smith}, {Smith}, {Sozzetti}, {Thuillot}, {van Reeven}, {Viala}, {Abbas}, {Abreu Aramburu}, {Accart}, {Aguado}, {Allan}, {Allasia}, {Altavilla}, {{\'A}lvarez}, {Alves}, {Anderson}, {Andrei}, {Anglada Varela}, {Antiche}, {Antoja}, {Ant{\'o}n}, {Arcay}, {Atzei}, {Ayache}, {Bach}, {Baker}, {Balaguer-N{\'u}{\~n}ez}, {Barache}, {Barata}, {Barbier}, {Barblan}, {Baroni}, {Barrado y Navascu{\'e}s}, {Barros}, {Barstow}, {Becciani}, {Bellazzini}, {Bellei}, {Bello Garc{\'\i}a}, {Belokurov}, {Bendjoya}, {Berihuete}, {Bianchi}, {Bienaym{\'e}}, {Billebaud}, {Blagorodnova}, {Blanco-Cuaresma}, {Boch},
  {Bombrun}, {Borrachero}, {Bouquillon}, {Bourda}, {Bouy}, {Bragaglia}, {Breddels}, {Brouillet}, {Br{\"u}semeister}, {Bucciarelli}, {Budnik}, {Burgess}, {Burgon}, {Burlacu}, {Busonero}, {Buzzi}, {Caffau}, {Cambras}, {Campbell}, {Cancelliere}, {Cantat-Gaudin}, {Carlucci}, {Carrasco}, {Castellani}, {Charlot}, {Charnas}, {Charvet}, {Chassat}, {Chiavassa}, {Clotet}, {Cocozza}, {Collins}, {Collins}, {Costigan}, {Crifo}, {Cross}, {Crosta}, {Crowley}, {Dafonte}, {Damerdji}, {Dapergolas}, {David}, {David}, {De Cat}, {de Felice}, {de Laverny}, {De Luise}, {De March}, {de Martino}, {de Souza}, {Debosscher}, {del Pozo}, {Delbo}, {Delgado}, {Delgado}, {di Marco}, {Di Matteo}, {Diakite}, {Distefano}, {Dolding}, {Dos Anjos}, {Drazinos}, {Dur{\'a}n}, {Dzigan}, {Ecale}, {Edvardsson}, {Enke}, {Erdmann}, {Escolar}, {Espina}, {Evans}, {Eynard Bontemps}, {Fabre}, {Fabrizio}, {Faigler}, {Falc{\~a}o}, {Farr{\`a}s Casas}, {Faye}, {Federici}, {Fedorets}, {Fern{\'a}ndez-Hern{\'a}ndez}, {Fernique}, {Fienga}, {Figueras}, {Filippi},
  {Findeisen}, {Fonti}, {Fouesneau}, {Fraile}, {Fraser}, {Fuchs}, {Furnell}, {Gai}, {Galleti}, {Galluccio}, {Garabato}, {Garc{\'\i}a-Sedano}, {Gar{\'e}}, {Garofalo}, {Garralda}, {Gavras}, {Gerssen}, {Geyer}, {Gilmore}, {Girona}, {Giuffrida}, {Gomes}, {Gonz{\'a}lez-Marcos}, {Gonz{\'a}lez-N{\'u}{\~n}ez}, {Gonz{\'a}lez-Vidal}, {Granvik}, {Guerrier}, {Guillout}, {Guiraud}, {G{\'u}rpide}, {Guti{\'e}rrez-S{\'a}nchez}, {Guy}, {Haigron}, {Hatzidimitriou}, {Haywood}, {Heiter}, {Helmi}, {Hobbs}, {Hofmann}, {Holl}, {Holland}, {Hunt}, {Hypki}, {Icardi}, {Irwin}, {Jevardat de Fombelle}, {Jofr{\'e}}, {Jonker}, {Jorissen}, {Julbe}, {Karampelas}, {Kochoska}, {Kohley}, {Kolenberg}, {Kontizas}, {Koposov}, {Kordopatis}, {Koubsky}, {Kowalczyk}, {Krone-Martins}, {Kudryashova}, {Kull}, {Bachchan}, {Lacoste-Seris}, {Lanza}, {Lavigne}, {Le Poncin-Lafitte}, {Lebreton}, {Lebzelter}, {Leccia}, {Leclerc}, {Lecoeur-Taibi}, {Lemaitre}, {Lenhardt}, {Leroux}, {Liao}, {Licata}, {Lindstr{\o}m}, {Lister}, {Livanou}, {Lobel}, {L{\"o}ffler},
  {L{\'o}pez}, {Lopez-Lozano}, {Lorenz}, {Loureiro}, {MacDonald}, {Magalh{\~a}es Fernandes}, {Managau}, {Mann}, {Mantelet}, {Marchal}, {Marchant}, {Marconi}, {Marie}, {Marinoni}, {Marrese}, {Marschalk{\'o}}, {Marshall}, {Mart{\'\i}n-Fleitas}, {Martino}, {Mary}, {Matijevi{\v{c}}}, {Mazeh}, {McMillan}, {Messina}, {Mestre}, {Michalik}, {Millar}, {Miranda}, {Molina}, {Molinaro}, {Molinaro}, {Moln{\'a}r}, {Moniez}, {Montegriffo}, {Monteiro}, {Mor}, {Mora}, {Morbidelli}, {Morel}, {Morgenthaler}, {Morley}, {Morris}, {Mulone}, {Muraveva}, {Musella}, {Narbonne}, {Nelemans}, {Nicastro}, {Noval}, {Ord{\'e}novic}, {Ordieres-Mer{\'e}}, {Osborne}, {Pagani}, {Pagano}, {Pailler}, {Palacin}, {Palaversa}, {Parsons}, {Paulsen}, {Pecoraro}, {Pedrosa}, {Pentik{\"a}inen}, {Pereira}, {Pichon}, {Piersimoni}, {Pineau}, {Plachy}, {Plum}, {Poujoulet}, {Pr{\v{s}}a}, {Pulone}, {Ragaini}, {Rago}, {Rambaux}, {Ramos-Lerate}, {Ranalli}, {Rauw}, {Read}, {Regibo}, {Renk}, {Reyl{\'e}}, {Ribeiro}, {Rimoldini}, {Ripepi}, {Riva}, {Rixon},
  {Roelens}, {Romero-G{\'o}mez}, {Rowell}, {Royer}, {Rudolph}, {Ruiz-Dern}, {Sadowski}, {Sagrist{\`a} Sell{\'e}s}, {Sahlmann}, {Salgado}, {Salguero}, {Sarasso}, {Savietto}, {Schnorhk}, {Schultheis}, {Sciacca}, {Segol}, {Segovia}, {Segransan}, {Serpell}, {Shih}, {Smareglia}, {Smart}, {Smith}, {Solano}, {Solitro}, {Sordo}, {Soria Nieto}, {Souchay}, {Spagna}, {Spoto}, {Stampa}, {Steele}, {Steidelm{\"u}ller}, {Stephenson}, {Stoev}, {Suess}, {S{\"u}veges}, {Surdej}, {Szabados}, {Szegedi-Elek}, {Tapiador}, {Taris}, {Tauran}, {Taylor}, {Teixeira}, {Terrett}, {Tingley}, {Trager}, {Turon}, {Ulla}, {Utrilla}, {Valentini}, {van Elteren}, {Van Hemelryck}, {van Leeuwen}, {Varadi}, {Vecchiato}, {Veljanoski}, {Via}, {Vicente}, {Vogt}, {Voss}, {Votruba}, {Voutsinas}, {Walmsley}, {Weiler}, {Weingrill}, {Werner}, {Wevers}, {Whitehead}, {Wyrzykowski}, {Yoldas}, {{\v{Z}}erjal}, {Zucker}, {Zurbach}, {Zwitter}, {Alecu}, {Allen}, {Allende Prieto}, {Amorim}, {Anglada-Escud{\'e}}, {Arsenijevic}, {Azaz}, {Balm}, {Beck}, {Bernstein},
  {Bigot}, {Bijaoui}, {Blasco}, {Bonfigli}, {Bono}, {Boudreault}, {Bressan}, {Brown}, {Brunet}, {Bunclark}, {Buonanno}, {Butkevich}, {Carret}, {Carrion}, {Chemin}, {Ch{\'e}reau}, {Corcione}, {Darmigny}, {de Boer}, {de Teodoro}, {de Zeeuw}, {Delle Luche}, {Domingues}, {Dubath}, {Fodor}, {Fr{\'e}zouls}, {Fries}, {Fustes}, {Fyfe}, {Gallardo}, {Gallegos}, {Gardiol}, {Gebran}, {Gomboc}, {G{\'o}mez}, {Grux}, {Gueguen}, {Heyrovsky}, {Hoar}, {Iannicola}, {Isasi Parache}, {Janotto}, {Joliet}, {Jonckheere}, {Keil}, {Kim}, {Klagyivik}, {Klar}, {Knude}, {Kochukhov}, {Kolka}, {Kos}, {Kutka}, {Lainey}, {LeBouquin}, {Liu}, {Loreggia}, {Makarov}, {Marseille}, {Martayan}, {Martinez-Rubi}, {Massart}, {Meynadier}, {Mignot}, {Munari}, {Nguyen}, {Nordlander}, {Ocvirk}, {O'Flaherty}, {Olias Sanz}, {Ortiz}, {Osorio}, {Oszkiewicz}, {Ouzounis}, {Palmer}, {Park}, {Pasquato}, {Peltzer}, {Peralta}, {P{\'e}turaud}, {Pieniluoma}, {Pigozzi}, {Poels}, {Prat}, {Prod'homme}, {Raison}, {Rebordao}, {Risquez}, {Rocca-Volmerange}, {Rosen},
  {Ruiz-Fuertes}, {Russo}, {Sembay}, {Serraller Vizcaino}, {Short}, {Siebert}, {Silva}, {Sinachopoulos}, {Slezak}, {Soffel}, {Sosnowska}, {Strai{\v{z}}ys}, {ter Linden}, {Terrell}, {Theil}, {Tiede}, {Troisi}, {Tsalmantza}, {Tur}, {Vaccari}, {Vachier}, {Valles}, {Van Hamme}, {Veltz}, {Virtanen}, {Wallut}, {Wichmann}, {Wilkinson}, {Ziaeepour}, \& {Zschocke}}]{2016A&A...595A...1G}
{Gaia Collaboration}, {Prusti}, T., {de Bruijne}, J.~H.~J., {et~al.} 2016, \aap, 595, A1, \dodoi{10.1051/0004-6361/201629272}

\bibitem[{{Gaia Collaboration} {et~al.}(2023){Gaia Collaboration}, {Vallenari}, {Brown}, {Prusti}, {de Bruijne}, {Arenou}, {Babusiaux}, {Biermann}, {Creevey}, {Ducourant}, {Evans}, {Eyer}, {Guerra}, {Hutton}, {Jordi}, {Klioner}, {Lammers}, {Lindegren}, {Luri}, {Mignard}, {Panem}, {Pourbaix}, {Randich}, {Sartoretti}, {Soubiran}, {Tanga}, {Walton}, {Bailer-Jones}, {Bastian}, {Drimmel}, {Jansen}, {Katz}, {Lattanzi}, {van Leeuwen}, {Bakker}, {Cacciari}, {Casta{\~n}eda}, {De Angeli}, {Fabricius}, {Fouesneau}, {Fr{\'e}mat}, {Galluccio}, {Guerrier}, {Heiter}, {Masana}, {Messineo}, {Mowlavi}, {Nicolas}, {Nienartowicz}, {Pailler}, {Panuzzo}, {Riclet}, {Roux}, {Seabroke}, {Sordo}, {Th{\'e}venin}, {Gracia-Abril}, {Portell}, {Teyssier}, {Altmann}, {Andrae}, {Audard}, {Bellas-Velidis}, {Benson}, {Berthier}, {Blomme}, {Burgess}, {Busonero}, {Busso}, {C{\'a}novas}, {Carry}, {Cellino}, {Cheek}, {Clementini}, {Damerdji}, {Davidson}, {de Teodoro}, {Nu{\~n}ez Campos}, {Delchambre}, {Dell'Oro}, {Esquej},
  {Fern{\'a}ndez-Hern{\'a}ndez}, {Fraile}, {Garabato}, {Garc{\'\i}a-Lario}, {Gosset}, {Haigron}, {Halbwachs}, {Hambly}, {Harrison}, {Hern{\'a}ndez}, {Hestroffer}, {Hodgkin}, {Holl}, {Jan{\ss}en}, {Jevardat de Fombelle}, {Jordan}, {Krone-Martins}, {Lanzafame}, {L{\"o}ffler}, {Marchal}, {Marrese}, {Moitinho}, {Muinonen}, {Osborne}, {Pancino}, {Pauwels}, {Recio-Blanco}, {Reyl{\'e}}, {Riello}, {Rimoldini}, {Roegiers}, {Rybizki}, {Sarro}, {Siopis}, {Smith}, {Sozzetti}, {Utrilla}, {van Leeuwen}, {Abbas}, {{\'A}brah{\'a}m}, {Abreu Aramburu}, {Aerts}, {Aguado}, {Ajaj}, {Aldea-Montero}, {Altavilla}, {{\'A}lvarez}, {Alves}, {Anders}, {Anderson}, {Anglada Varela}, {Antoja}, {Baines}, {Baker}, {Balaguer-N{\'u}{\~n}ez}, {Balbinot}, {Balog}, {Barache}, {Barbato}, {Barros}, {Barstow}, {Bartolom{\'e}}, {Bassilana}, {Bauchet}, {Becciani}, {Bellazzini}, {Berihuete}, {Bernet}, {Bertone}, {Bianchi}, {Binnenfeld}, {Blanco-Cuaresma}, {Blazere}, {Boch}, {Bombrun}, {Bossini}, {Bouquillon}, {Bragaglia}, {Bramante}, {Breedt},
  {Bressan}, {Brouillet}, {Brugaletta}, {Bucciarelli}, {Burlacu}, {Butkevich}, {Buzzi}, {Caffau}, {Cancelliere}, {Cantat-Gaudin}, {Carballo}, {Carlucci}, {Carnerero}, {Carrasco}, {Casamiquela}, {Castellani}, {Castro-Ginard}, {Chaoul}, {Charlot}, {Chemin}, {Chiaramida}, {Chiavassa}, {Chornay}, {Comoretto}, {Contursi}, {Cooper}, {Cornez}, {Cowell}, {Crifo}, {Cropper}, {Crosta}, {Crowley}, {Dafonte}, {Dapergolas}, {David}, {David}, {de Laverny}, {De Luise}, {De March}, {De Ridder}, {de Souza}, {de Torres}, {del Peloso}, {del Pozo}, {Delbo}, {Delgado}, {Delisle}, {Demouchy}, {Dharmawardena}, {Di Matteo}, {Diakite}, {Diener}, {Distefano}, {Dolding}, {Edvardsson}, {Enke}, {Fabre}, {Fabrizio}, {Faigler}, {Fedorets}, {Fernique}, {Fienga}, {Figueras}, {Fournier}, {Fouron}, {Fragkoudi}, {Gai}, {Garcia-Gutierrez}, {Garcia-Reinaldos}, {Garc{\'\i}a-Torres}, {Garofalo}, {Gavel}, {Gavras}, {Gerlach}, {Geyer}, {Giacobbe}, {Gilmore}, {Girona}, {Giuffrida}, {Gomel}, {Gomez}, {Gonz{\'a}lez-N{\'u}{\~n}ez},
  {Gonz{\'a}lez-Santamar{\'\i}a}, {Gonz{\'a}lez-Vidal}, {Granvik}, {Guillout}, {Guiraud}, {Guti{\'e}rrez-S{\'a}nchez}, {Guy}, {Hatzidimitriou}, {Hauser}, {Haywood}, {Helmer}, {Helmi}, {Sarmiento}, {Hidalgo}, {Hilger}, {H{\l}adczuk}, {Hobbs}, {Holland}, {Huckle}, {Jardine}, {Jasniewicz}, {Jean-Antoine Piccolo}, {Jim{\'e}nez-Arranz}, {Jorissen}, {Juaristi Campillo}, {Julbe}, {Karbevska}, {Kervella}, {Khanna}, {Kontizas}, {Kordopatis}, {Korn}, {K{\'o}sp{\'a}l}, {Kostrzewa-Rutkowska}, {Kruszy{\'n}ska}, {Kun}, {Laizeau}, {Lambert}, {Lanza}, {Lasne}, {Le Campion}, {Lebreton}, {Lebzelter}, {Leccia}, {Leclerc}, {Lecoeur-Taibi}, {Liao}, {Licata}, {Lindstr{\o}m}, {Lister}, {Livanou}, {Lobel}, {Lorca}, {Loup}, {Madrero Pardo}, {Magdaleno Romeo}, {Managau}, {Mann}, {Manteiga}, {Marchant}, {Marconi}, {Marcos}, {Marcos Santos}, {Mar{\'\i}n Pina}, {Marinoni}, {Marocco}, {Marshall}, {Martin Polo}, {Mart{\'\i}n-Fleitas}, {Marton}, {Mary}, {Masip}, {Massari}, {Mastrobuono-Battisti}, {Mazeh}, {McMillan}, {Messina}, {Michalik},
  {Millar}, {Mints}, {Molina}, {Molinaro}, {Moln{\'a}r}, {Monari}, {Mongui{\'o}}, {Montegriffo}, {Montero}, {Mor}, {Mora}, {Morbidelli}, {Morel}, {Morris}, {Muraveva}, {Murphy}, {Musella}, {Nagy}, {Noval}, {Oca{\~n}a}, {Ogden}, {Ordenovic}, {Osinde}, {Pagani}, {Pagano}, {Palaversa}, {Palicio}, {Pallas-Quintela}, {Panahi}, {Payne-Wardenaar}, {Pe{\~n}alosa Esteller}, {Penttil{\"a}}, {Pichon}, {Piersimoni}, {Pineau}, {Plachy}, {Plum}, {Poggio}, {Pr{\v{s}}a}, {Pulone}, {Racero}, {Ragaini}, {Rainer}, {Raiteri}, {Rambaux}, {Ramos}, {Ramos-Lerate}, {Re Fiorentin}, {Regibo}, {Richards}, {Rios Diaz}, {Ripepi}, {Riva}, {Rix}, {Rixon}, {Robichon}, {Robin}, {Robin}, {Roelens}, {Rogues}, {Rohrbasser}, {Romero-G{\'o}mez}, {Rowell}, {Royer}, {Ruz Mieres}, {Rybicki}, {Sadowski}, {S{\'a}ez N{\'u}{\~n}ez}, {Sagrist{\`a} Sell{\'e}s}, {Sahlmann}, {Salguero}, {Samaras}, {Sanchez Gimenez}, {Sanna}, {Santove{\~n}a}, {Sarasso}, {Schultheis}, {Sciacca}, {Segol}, {Segovia}, {S{\'e}gransan}, {Semeux}, {Shahaf}, {Siddiqui}, {Siebert},
  {Siltala}, {Silvelo}, {Slezak}, {Slezak}, {Smart}, {Snaith}, {Solano}, {Solitro}, {Souami}, {Souchay}, {Spagna}, {Spina}, {Spoto}, {Steele}, {Steidelm{\"u}ller}, {Stephenson}, {S{\"u}veges}, {Surdej}, {Szabados}, {Szegedi-Elek}, {Taris}, {Taylor}, {Teixeira}, {Tolomei}, {Tonello}, {Torra}, {Torra}, {Torralba Elipe}, {Trabucchi}, {Tsounis}, {Turon}, {Ulla}, {Unger}, {Vaillant}, {van Dillen}, {van Reeven}, {Vanel}, {Vecchiato}, {Viala}, {Vicente}, {Voutsinas}, {Weiler}, {Wevers}, {Wyrzykowski}, {Yoldas}, {Yvard}, {Zhao}, {Zorec}, {Zucker}, \& {Zwitter}}]{2023A&A...674A...1G}
{Gaia Collaboration}, {Vallenari}, A., {Brown}, A.~G.~A., {et~al.} 2023, \aap, 674, A1, \dodoi{10.1051/0004-6361/202243940}

\bibitem[{{Hotan} {et~al.}(2004){Hotan}, {van Straten}, \& {Manchester}}]{2004PASA...21..302H}
{Hotan}, A.~W., {van Straten}, W., \& {Manchester}, R.~N. 2004, \pasa, 21, 302, \dodoi{10.1071/AS04022}

\bibitem[{{Istrate} {et~al.}(2016){Istrate}, {Marchant}, {Tauris}, {Langer}, {Stancliffe}, \& {Grassitelli}}]{2016A&A...595A..35I}
{Istrate}, A.~G., {Marchant}, P., {Tauris}, T.~M., {et~al.} 2016, \aap, 595, A35, \dodoi{10.1051/0004-6361/201628874}

\bibitem[{{Jeans}(1924)}]{1924MNRAS..85....2J}
{Jeans}, J.~H. 1924, \mnras, 85, 2, \dodoi{10.1093/mnras/85.1.2}

\bibitem[{{Jeans}(1928)}]{1928asco.book.....J}
---. 1928, {Astronomy and cosmogony}

\bibitem[{{Khechinashvili} {et~al.}(2000){Khechinashvili}, {Melikidze}, \& {Gil}}]{2000ApJ...541..335K}
{Khechinashvili}, D.~G., {Melikidze}, G.~I., \& {Gil}, J.~A. 2000, \apj, 541, 335, \dodoi{10.1086/309408}

\bibitem[{{Kirichenko} {et~al.}(2024){Kirichenko}, {Zharikov}, {Karpova}, {Fonseca}, {Zyuzin}, {Shibanov}, {L{\'o}pez}, {Gilfanov}, {Cabrera-Lavers}, {Geier}, {Dong}, {Good}, {McKee}, {Meyers}, {Stairs}, {McLaughlin}, \& {Swiggum}}]{kzk+24}
{Kirichenko}, A.~Y., {Zharikov}, S.~V., {Karpova}, A.~V., {et~al.} 2024, Monthly Notices of the Royal Astronomical Society, 527, 4563, \dodoi{10.1093/mnras/stad3391}

\bibitem[{{Kosakowski} {et~al.}(2023){Kosakowski}, {Brown}, {Kilic}, {Kupfer}, {B{\'e}dard}, {Gianninas}, {Ag{\"u}eros}, \& {Barrientos}}]{2023ApJ...950..141K}
{Kosakowski}, A., {Brown}, W.~R., {Kilic}, M., {et~al.} 2023, \apj, 950, 141, \dodoi{10.3847/1538-4357/acd187}

\bibitem[{{Kuijken} \& {Gilmore}(1989{\natexlab{a}})}]{1989MNRAS.239..571K}
{Kuijken}, K., \& {Gilmore}, G. 1989{\natexlab{a}}, \mnras, 239, 571, \dodoi{10.1093/mnras/239.2.571}

\bibitem[{{Kuijken} \& {Gilmore}(1989{\natexlab{b}})}]{1989MNRAS.239..605K}
---. 1989{\natexlab{b}}, \mnras, 239, 605, \dodoi{10.1093/mnras/239.2.605}

\bibitem[{Kumar {et~al.}(2023)}]{MUSES:2023hyz}
Kumar, R., {et~al.} 2023.
\newblock \doarXiv{2303.17021}

\bibitem[{Lattimer \& Prakash(2001)}]{Lattimer:2000nx}
Lattimer, J.~M., \& Prakash, M. 2001, Astrophys. J., 550, 426, \dodoi{10.1086/319702}

\bibitem[{{Lazaridis} {et~al.}(2009){Lazaridis}, {Wex}, {Jessner}, {Kramer}, {Stappers}, {Janssen}, {Desvignes}, {Purver}, {Cognard}, {Theureau}, {Lyne}, {Jordan}, \& {Zensus}}]{2009MNRAS.400..805L}
{Lazaridis}, K., {Wex}, N., {Jessner}, A., {et~al.} 2009, \mnras, 400, 805, \dodoi{10.1111/j.1365-2966.2009.15481.x}

\bibitem[{Linares {et~al.}(2018)Linares, Shahbaz, \& Casares}]{Linares:2018ppq}
Linares, M., Shahbaz, T., \& Casares, J. 2018, Astrophys. J., 859, 54, \dodoi{10.3847/1538-4357/aabde6}

\bibitem[{{Lorimer} \& {Kramer}(2004)}]{2004hpa..book.....L}
{Lorimer}, D.~R., \& {Kramer}, M. 2004, {Handbook of Pulsar Astronomy}, Vol.~4

\bibitem[{{Luo} {et~al.}(2019){Luo}, {Ransom}, {Demorest}, {van Haasteren}, {Ray}, {Stovall}, {Bachetti}, {Archibald}, {Kerr}, {Colen}, \& {Jenet}}]{2019ascl.soft02007L}
{Luo}, J., {Ransom}, S., {Demorest}, P., {et~al.} 2019, {PINT: High-precision pulsar timing analysis package}, Astrophysics Source Code Library, record ascl:1902.007

\bibitem[{Luo {et~al.}(2021)}]{Luo:2020ksx}
Luo, J., {et~al.} 2021, Astrophys. J., 911, 45, \dodoi{10.3847/1538-4357/abe62f}

\bibitem[{Lynch {et~al.}(2013)}]{Lynch:2012vv}
Lynch, R.~S., {et~al.} 2013, Astrophys. J., 763, 81, \dodoi{10.1088/0004-637X/763/2/81}

\bibitem[{Mamajek {et~al.}(2015)}]{IAUInter-DivisionA-GWorkingGrouponNominalUnitsforStellarPlanetaryAstronomy:2015fjh}
Mamajek, E.~E., {et~al.} 2015.
\newblock \doarXiv{1510.07674}

\bibitem[{{Manchester} {et~al.}(2005){Manchester}, {Hobbs}, {Teoh}, \& {Hobbs}}]{atnfcatalog}
{Manchester}, R.~N., {Hobbs}, G.~B., {Teoh}, A., \& {Hobbs}, M. 2005, AJ, 129, 1993, \dodoi{10.1086/428488}

\bibitem[{{Moran} {et~al.}(2024){Moran}, {Mingarelli}, {Van Tilburg}, \& {Good}}]{2024PhRvD.109l3015M}
{Moran}, A., {Mingarelli}, C. M.~F., {Van Tilburg}, K., \& {Good}, D. 2024, \prd, 109, 123015, \dodoi{10.1103/PhysRevD.109.123015}

\bibitem[{{Morello} {et~al.}(2019){Morello}, {Barr}, {Cooper}, {Bailes}, {Bates}, {Bhat}, {Burgay}, {Burke-Spolaor}, {Cameron}, {Champion}, {Eatough}, {Flynn}, {Jameson}, {Johnston}, {Keith}, {Keane}, {Kramer}, {Levin}, {Ng}, {Petroff}, {Possenti}, {Stappers}, {van Straten}, \& {Tiburzi}}]{2019MNRAS.483.3673M}
{Morello}, V., {Barr}, E.~D., {Cooper}, S., {et~al.} 2019, \mnras, 483, 3673, \dodoi{10.1093/mnras/sty3328}

\bibitem[{{Morello} {et~al.}(2023){Morello}, {Barr}, {Cooper}, {Bailes}, {Bates}, {Bhat}, {Burgay}, {Burke-Spolaor}, {Cameron}, {Champion}, {Eatough}, {Flynn}, {Jameson}, {Johnston}, {Keith}, {Keane}, {Kramer}, {Levin}, {Ng}, {Petroff}, {Possenti}, {Stappers}, {van Straten}, \& {Tiburzi}}]{2023ascl.soft10008M}
---. 2023, {clfd: Clean folded data}, Astrophysics Source Code Library, record ascl:2310.008

\bibitem[{{Nice} \& {Taylor}(1995)}]{1995ApJ...441..429N}
{Nice}, D.~J., \& {Taylor}, J.~H. 1995, \apj, 441, 429, \dodoi{10.1086/175367}

\bibitem[{\"Ozel \& Freire(2016)}]{Ozel:2016oaf}
\"Ozel, F., \& Freire, P. 2016, Ann. Rev. Astron. Astrophys., 54, 401, \dodoi{10.1146/annurev-astro-081915-023322}

\bibitem[{Park {et~al.}(2021)Park, Folkner, Williams, \& Boggs}]{Park_2021}
Park, R.~S., Folkner, W.~M., Williams, J.~G., \& Boggs, D.~H. 2021, The Astronomical Journal, 161, 105, \dodoi{https://doi.org/10.3847/1538-3881/abd414}

\bibitem[{Pennucci(2019)}]{Pennucci:2018zow}
Pennucci, T.~T. 2019, Astrophys. J., 871, 34, \dodoi{10.3847/1538-4357/aaf6ef}

\bibitem[{Pennucci {et~al.}(2014)Pennucci, Demorest, \& Ransom}]{Pennucci:2014dja}
Pennucci, T.~T., Demorest, P.~B., \& Ransom, S.~M. 2014, Astrophys. J., 790, 93, \dodoi{10.1088/0004-637X/790/2/93}

\bibitem[{{Phinney}(1992)}]{1992RSPTA.341...39P}
{Phinney}, E.~S. 1992, Philosophical Transactions of the Royal Society of London Series A, 341, 39, \dodoi{10.1098/rsta.1992.0084}

\bibitem[{Possenti {et~al.}(2003)Possenti, D'Amico, Manchester, Camilo, Lyne, Sarkissian, \& Corongiu}]{Possenti:2003nr}
Possenti, A., D'Amico, N., Manchester, R.~N., {et~al.} 2003, Astrophys. J., 599, 475, \dodoi{10.1086/379190}

\bibitem[{{Ransom}(2011)}]{2011ascl.soft07017R}
{Ransom}, S. 2011, {PRESTO: PulsaR Exploration and Search TOolkit}, Astrophysics Source Code Library, record ascl:1107.017

\bibitem[{Ransom {et~al.}(2004)Ransom, Stairs, Backer, Greenhill, Hessels, \& Kaspi}]{Ransom:2003qv}
Ransom, S.~M., Stairs, I.~H., Backer, D.~C., {et~al.} 2004, Astrophys. J., 604, 328, \dodoi{10.1086/381730}

\bibitem[{{Reid} {et~al.}(2014){Reid}, {Menten}, {Brunthaler}, {Zheng}, {Dame}, {Xu}, {Wu}, {Zhang}, {Sanna}, {Sato}, {Hachisuka}, {Choi}, {Immer}, {Moscadelli}, {Rygl}, \& {Bartkiewicz}}]{2014ApJ...783..130R}
{Reid}, M.~J., {Menten}, K.~M., {Brunthaler}, A., {et~al.} 2014, \apj, 783, 130, \dodoi{10.1088/0004-637X/783/2/130}

\bibitem[{Romani {et~al.}(2021)Romani, Kandel, Filippenko, Brink, \& Zheng}]{Romani:2021xmb}
Romani, R.~W., Kandel, D., Filippenko, A.~V., Brink, T.~G., \& Zheng, W. 2021, Astrophys. J. Lett., 908, L46, \dodoi{10.3847/2041-8213/abe2b4}

\bibitem[{Romani {et~al.}(2022)Romani, Kandel, Filippenko, Brink, \& Zheng}]{Romani:2022jhd}
---. 2022, Astrophys. J. Lett., 934, L17, \dodoi{10.3847/2041-8213/ac8007}

\bibitem[{Sen {et~al.}(2024)Sen, Linares, Kennedy, Breton, Misra, Turchetta, Dhillon, Sanchez, \& Clark}]{Sen:2024xbs}
Sen, B., Linares, M., Kennedy, M.~R., {et~al.} 2024.
\newblock \doarXiv{2407.10800}

\bibitem[{Shao \& Yagi(2022)}]{Shao:2022koz}
Shao, L., \& Yagi, K. 2022, Sci. Bull., 67, 1946, \dodoi{10.1016/j.scib.2022.09.018}

\bibitem[{{Shapiro}(1964)}]{Shapiro64}
{Shapiro}, I.~I. 1964, Phys. Rev. Lett., 13, 789, \dodoi{10.1103/PhysRevLett.13.789}

\bibitem[{{Shklovskii}(1970)}]{1970SvA....13..562S}
{Shklovskii}, I.~S. 1970, \sovast, 13, 562

\bibitem[{{Smarr} \& {Blandford}(1976)}]{1976ApJ...207..574S}
{Smarr}, L.~L., \& {Blandford}, R. 1976, \apj, 207, 574, \dodoi{10.1086/154524}

\bibitem[{{Stairs}(2003)}]{Stairs:2003}
{Stairs}, I.~H. 2003, Living Reviews in Relativity, 6, 5, \dodoi{10.12942/lrr-2003-5}

\bibitem[{Susobhanan {et~al.}(2024)}]{Susobhanan:2024gzf}
Susobhanan, A., {et~al.} 2024.
\newblock \doarXiv{2405.01977}

\bibitem[{{Taylor}(1992)}]{1992RSPTA.341..117T}
{Taylor}, J.~H. 1992, Philosophical Transactions of the Royal Society of London Series A, 341, 117, \dodoi{10.1098/rsta.1992.0088}

\bibitem[{{Taylor} \& {Weisberg}(1989)}]{1989ApJ...345..434T}
{Taylor}, J.~H., \& {Weisberg}, J.~M. 1989, \apj, 345, 434, \dodoi{10.1086/167917}

\bibitem[{{van Straten} \& {Bailes}(2011)}]{vb11}
{van Straten}, W., \& {Bailes}, M. 2011, PASA, 28, 1, \dodoi{10.1071/AS10021}

\bibitem[{{van Straten} {et~al.}(2011){van Straten}, {Demorest}, {Khoo}, {Keith}, {Hotan}, \& {et al.}}]{2011ascl.soft05014V}
{van Straten}, W., {Demorest}, P., {Khoo}, J., {et~al.} 2011, {PSRCHIVE: Development Library for the Analysis of Pulsar Astronomical Data}, Astrophysics Source Code Library, record ascl:1105.014

\bibitem[{{van Straten} {et~al.}(2012){van Straten}, {Demorest}, \& {Oslowski}}]{2012AR&T....9..237V}
{van Straten}, W., {Demorest}, P., \& {Oslowski}, S. 2012, Astronomical Research and Technology, 9, 237, \dodoi{10.48550/arXiv.1205.6276}

\bibitem[{{van Straten} {et~al.}(2010){van Straten}, {Manchester}, {Johnston}, \& {Reynolds}}]{2010PASA...27..104V}
{van Straten}, W., {Manchester}, R.~N., {Johnston}, S., \& {Reynolds}, J.~E. 2010, \pasa, 27, 104, \dodoi{10.1071/AS09084}

\bibitem[{Will(1993)}]{Will:1993hxu}
Will, C.~M. 1993, {Theory and Experiment in Gravitational Physics}, \dodoi{10.1017/CBO9780511564246}

\end{thebibliography}
\bibliographystyle{aasjournal}

\end{document}